\documentclass[prx,nofootinbib]{revtex4}

\usepackage{graphicx}
\usepackage{ulem} % Pacote a ser removido da versão final
\usepackage{amsmath,srcltx}
\usepackage{latexsym}
\usepackage{amssymb}    
\usepackage{amsfonts}
\usepackage{tikz}
\usepackage{float}
\usepackage{slashed}
\usepackage{caption}
\usepackage{array}

\newcommand{\be}{\begin{equation}}
\newcommand{\ee}{\end{equation}}
\newcommand{\ba}{\begin{eqnarray}}
\newcommand{\ea}{\end{eqnarray}}

\def\a{\alpha}
\def\b{\beta}

\def\ve{\varepsilon}

\def\g{\gamma}

\def\m{\mu}
\def\n{\nu}
\def\o{\omega}

\def\r{\rho}
\def\s{\sigma}

\def\D{\Delta}

\def\O{\Omega}

\def\S{\Sigma}

\def\ca{{\cal A}}

\newcommand{\pa}{\partial}

\def\sl#1{\rlap{\hbox{$\mskip 1 mu /$}}#1}

\def\I{\leavevmode\hbox{\small1\kern-3.8pt\normalsize1}}

\begin{document}

\title{The spectrum consistency of fractional quantum Hall effect model}

\author{O.M. Del Cima} 
\email{oswaldo.delcima@ufv.br} 
\affiliation{Universidade Federal de Vi\c cosa (UFV),\\
Departamento de F\'\i sica - Campus Universit\'ario,\\
Avenida Peter Henry Rolfs s/n - 36570-900 - Vi\c cosa - MG - Brazil.}
\affiliation{Ibitipoca Institute of Physics (IbitiPhys),\\
36140-000 - Concei\c c\~ao do Ibitipoca - MG - Brazil.}

\author{L.S. Lima} 
\email{lazaro.lima@ufv.br}
\affiliation{Universidade Federal de Vi\c cosa (UFV),\\
Departamento de F\'\i sica - Campus Universit\'ario,\\
Avenida Peter Henry Rolfs s/n - 36570-900 - Vi\c cosa - MG - Brazil.}
\affiliation{Ibitipoca Institute of Physics (IbitiPhys),\\
36140-000 - Concei\c c\~ao do Ibitipoca - MG - Brazil.}

\author{E.S. Miranda} 
\email{emerson.s.miranda@ufv.br} 
\affiliation{Universidade Federal de Vi\c cosa (UFV),\\
Departamento de F\'\i sica - Campus Universit\'ario,\\
Avenida Peter Henry Rolfs s/n - 36570-900 - Vi\c cosa - MG - Brazil.}
\affiliation{Ibitipoca Institute of Physics (IbitiPhys),\\
36140-000 - Concei\c c\~ao do Ibitipoca - MG - Brazil.}

%===================================================================
\begin{abstract}
The spectrum consistency of the three fermion family $U(1)\times U(1)$ quantum electrodynamics in three space-time dimensions  is analyzed. It has been verified that the originally proposed action violates the unitarity consistency condition in the gauge field sector by exhibiting negative norm states in the spectrum. However, the cure came through the fundamental gauge fields properly settled, consequently, the action rewritten in terms of those fundamental fields shows to be free from any spurious degrees of freedom, and the model now becomes safe for further quantization.
\end{abstract}
%\pacs{11.10.Gh 11.15.-q 11.15.Bt 11.15.Ex}
\maketitle
%===================================================================

\textit{Introduction}---. The quantum electrodynamics in three space-time dimensions (QED$_3$) has drawn attention, since the early eighties thanks to the pioneering works by Deser, Jackiw, Templeton and Schonfeld \cite{deser-jackiw-templeton-schonfeld}, in view of its potentiality as a theoretical foundation for planar condensed matter phenomena, such as high-$T_{\rm c}$ superconductors \cite{high-Tc}, quantum Hall effect \cite{quantum-hall-effect}, topological insulators \cite{topological-insulators}, topological superconductors \cite{topological-superconductors} and graphene \cite{graphene}. Since then, the quantum electrodynamics in two space dimensions has been widely studied in many physical configurations, namely, fermions families, Abelian and non-Abelian gauge groups, small and large gauge transformations, compact space-times, space-times with boundaries, curved space-times, even and odd under parity, discrete space-times, external fields and finite temperatures. It is well established that lattice quantum field theories might be used to describe topological materials, which shows a correspondence between many phenomena in condensed matter physics and relativistic quantum field theory \cite{kaplan_2000}. Although the existence of many works in this sense, more investigations were needed, for instance in the context of fractional quantum Hall effect. Consequently in \cite{kaplan_2020}, Kaplan and Sen have proposed in three space-time dimensions, a three fermion family Lorentz invariant quantum electrodynamics with $U(1)\times U(1)$-local symmetry, exhibiting two massive gauge bosons. Moreover there, in the infrared limit, they obtained the fractional Hall conductivity as a function of the fermion flavor numbers.

In quantum field theory, causality and unitarity are elementary physical requirements. The causality establishes a time correlation between the cause and its subsequent effect, a principle expressing that any change in the interaction law in a space-time region can affect the evolution of the system only at subsequent times. On the other hand, the unitarity of the $S$-matrix reflects the fundamental principle of probability conservation, which means the absence of negative norm states in the spectrum. Although, in certain circumstances, the artificial procedure of an indefinite metric in Hilbert space has to be introduced, the physical quantities preserved through time evolution refer to positive-norm states \cite{bogoliubov_shirkov}. Spectrum consistency analysis shall be performed \cite{nieuwenhuizen}, prior to quantization, so as to avoid further obstruction in the renormalization procedure. In addition, for instance, within the Becchi-Rouet-Stora algebraic renormalization method \cite{piguet_sorella}, previous spectral analysis of any model should be realized, which consists of obtaining the propagators and analyzing the spectrum consistency at the tree-level, namely unitarity, and causality, and only after that, deeper analyses -- such as the fulfillment or not of the Slavnov-Taylor identity, thus the absence or not of gauge anomaly -- are performed, as in the proof of the quantum scale invariance in a parity-preserving graphene-like QED$_3$ \cite{del_cima_lima_miranda}. The lack of spectrum consistency at the tree level signals problems, it has been revealed, in three-dimensional space-time $U(1)$ Chern-Simons-Maxwell-Proca charged vector field model, where unitarity is explicitly jeopardized due to the violation of Froissart-Martin cross-section bound \cite{delcima_1994}. Naively, there would be no evident reason to expect that the model proposed in \cite{kaplan_2020} could face at the tree-level any problems concerning causality and unitarity indeed. Nevertheless, by analyzing the poles structure complexity of the vector fields propagators, where all poles depend on three fermion flavors and two coupling constants, precaution about causality (from the signal of the poles) and unitarity (from the signal of the imaginary part of the residue of the transition amplitude at the poles) seems to be necessary, mainly because our posterior aim is to quantize the model by adopting the algebraic renormalization method. 
In this work, we analyze the three fermion family QED$_3$ \cite{kaplan_2020} spectrum at the tree-level, {\it i.e.} causality, unitarity and quantum propagating degrees of freedom. In the meantime of this analysis, it has been identified that propagating ghosts (negative norm states) be present in the bosonic sector of the spectrum, possibly indicating that both vector fields from the proposed model \cite{kaplan_2020} were not the fundamental ones. Solving this issue, by determining the fundamental vector fields, we prove that the model is free from tachyons and ghosts, regardless any condition fermion flavors and coupling constants have to be fulfilled, consequently no spurious degrees of freedom propagate. However, this constraintless parameter spectrum behavior is accomplished only by the fundamental vector fields, which have been identified and the model rewritten in terms of them. Additionally, we determine the true photon field, which is much lighter
than that proposed in the effective photon action in \cite{kaplan_2020}. All details about the issues mentioned above shall be presented in the following.

\textit{The model}---. The three fermion family $U(1)\times U(1)$ QED$_3$ \cite{kaplan_2020} exhibits two massive gauge bosons, $A_\m$ and $Z_\m$, and three massive fermions, $\psi_i$, $\chi_j$ and $\o_k$, with flavor number, $n_\psi$, $n_\chi$ and $n_\o$, respectively. The action reads: 
\begin{align}\label{action}
\nonumber
\S_{\rm KS}&=\int d^3x \bigg\{ -\frac{1}{4}F^{\m\n}F_{\m\n}-\frac{1}{4}Z^{\m\n} Z_{\m\n}+ \ve^{\m\n\r}
\left[(n_\psi+n_\o)e^2 A_\m\pa_\n A_\r+(n_\chi+n_\o)g^2 Z_\m\pa_\n Z_\r
+2n_\o e g A_\m\pa_\n Z_\r\right]+ \\
%%%%%%%%%%%%%%%%%%%%%%%%%%%%%%%%%%%%%%%%%%%%%%%%%%%%%%%%%%bosonic%%%%%%%%%%%
& +\frac{n_\psi}{|n_\psi|}m \bar{\psi}_i\psi_i
+\frac{n_\chi}{|n_\chi|}m \bar{\chi}_j\chi_j+\frac{n_\o}{|n_\o|}m \bar{\o}_k\o_k
%%%%%%%%%%%%%%%%%%%%%%%%%%%%%%%%%%%%%%%%%%%%%%%%%%%%%%%%%%fermionic%%%%%%%%%
+i\bar{\psi}_i\slashed{D}\psi_i+ 
i\bar{\chi}_j\slashed{D}\chi_j+
i\bar{\o}_k\slashed{D}\o_k
-\frac{1}{2\a}(\pa^\m A_\m)^2-\frac{1}{2\b}(\pa^\m Z_\m)^2 \bigg\}~,
%%%%%%%%%%%%%%%%%%%%%%%%%%%%%%%%%%%%%%%%%%%%%%%%%%%%%%%%%%interaction%%%%%%%
\end{align}
where $i$, $j$, $k$ range from 1 to $|n_\psi|$, $|n_\chi|$, $|n_\o|$, respectively. Also,   $q_A$ and $q_Z$ are the charge assignments for the fermion fields and their values are, 1 and 0 for $\psi_i$, 0 and 1 for $\chi_j$, and 1 and 1 for $\o_k$, with $e$ and $g$ being the gauge coupling (positive) constants and $m$ a mass parameter. The covariant derivative $\slashed D\equiv\slashed \pa +iq_A e \slashed A +iq_Z g \slashed Z$, and the $\g$-matrices $\g^\m\equiv(\s_z,-i\s_x,i\s_y)$. From the action (\ref{action}), its free gauge field piece can be written as: 
\begin{equation}\label{bosonicact2}
  \S_{AZ}= \int d^3x \bigg\{\frac{1}{2}A_\mu\bigg[\square \left(\Theta^{\mu\nu}+\frac{1}{\a}\O^{\m\n}\right)+E S^{\mu\nu} \bigg]A_\nu+\frac{1}{2}Z_\mu\bigg[\square \left(\Theta^{\mu\nu}+\frac{1}{\b}\O^{\m\n}\right)+G S^{\mu\nu} \bigg] Z_\nu 
+ A_\mu \left[H S^{\mu\nu}\right]Z_\nu\bigg\}~,
\end{equation}
with 
\begin{equation}\label{definitions}
E\equiv 2(n_\psi+n_\o)e^2~,~~G\equiv 2(n_\chi+n_\o)g^2~,~~H\equiv 2n_\o eg~,~~\Theta^{\mu\nu}=\eta^{\mu\nu}-\frac{\pa^\mu\pa^\nu}{\square}~,~~\O^{\m\n}=\frac{\pa^\m\pa^\n}{\square}~,~~S^{\mu\nu}=\ve^{\m\r\n}\pa_\r~,
\end{equation}
where the algebra among the operators, $\Theta^{\mu\nu}$, $\O^{\m\n}$ and $S^{\mu\nu}$,  are displayed in Table \ref{table}. 

\begin{table}[H]
\centering
\begin{tabular}[c]{|m{0.8cm}|m{0.8cm}|m{0.8cm}|m{0.8cm}|}
	\hline
 & $~~~\Theta$ & $~~~\O$ & $~~~S$ \\
 \hline
 $~~~\Theta$& $~~~\Theta$ & $~~~0$&~~~$S$ \\
 \hline
 $~~~\O$ &$~~~0$ &$~~~\O$ &$~~~0$ \\
 \hline
 $~~~S$&$~~~S$ &$~~~0$ &$-\square \Theta$\\
 \hline
 \end{tabular}
 \caption{Algebra of the operators $\Theta$, $\O$ and $S$.}\label{table}
  \end{table}

From the gauge field free action (\ref{bosonicact2}), the propagators in the momentum space are given by: 
\begin{equation}\label{prop1}
  \D_{\m\n}^{ZZ}(k)= i\left[\frac{k^2-E^2-H^2}{\mathcal{D}}\left(\eta_{\m\n}-\frac{k_\m k_\n}{k^2}\right)+\frac{\b}{k^2}\frac{k_\m k_\n}{k^2}+
  \frac{1}{k^2}\frac{G(k^2-E^2)+H^2E}{\mathcal{D}}i\ve_{\m\r\n}k^\r\right]~,
\end{equation}
\begin{equation}\label{prop2}
  \D_{\m\n}^{AA}(k)= i\left[\frac{k^2-G^2-H^2}{\mathcal{D}}\left(\eta_{\m\n}-\frac{k_\m k_\n}{k^2}\right)+\frac{\a}{k^2}\frac{k_\m k_\n}{k^2}+
  \frac{1}{k^2}\frac{E(k^2-G^2)+H^2G}{\mathcal{D}}i\ve_{\m\r\n}k^\r\right]~,
\end{equation}
\begin{equation}\label{prop3}
  \D_{\m\n}^{AZ}(k)=\D_{\m\n}^{ZA}(k) = \frac{iH}{\mathcal{D}}\left[
    \frac{i\ve^{\m\r\n}k_\r}{k^2} (k^2 +GE-H^2)+\left(\eta_{\m\n}-\frac{k_\m k_\n}{k^2}\right)(E+G)
  \right]~,
\end{equation}
\begin{equation}\label{prop4}
  \D^{\phi}(k)\!= i\frac{{\sl k}-{n_\phi}/{|n_\phi|}m }{k^2-m^2}~,
\end{equation}
where $\phi$, in (\ref{prop4}), can be any of the fermion fields ($\psi_i$, $\chi_j$ and $\o_k$) and $\mathcal{D}\equiv (k^2-G^2)(k^2-E^2)-2H^2(k^2+EG)+H^4$.

\textit{The spectrum analysis}---. Owing to the fact that poles in the Feynman propagators are the particle masses, it is essential to investigate the former concerning causality and unitarity. Causality reflects the fact that the effects cannot precede the causes, and there is no instantaneous transmission of information, thus a change in any region of space-time only influences the system in posterior times. A model is causal if its spectrum is free from tachyons, particles with negative mass squares ($k^2<0$) in space-time metric signature $\eta_{\mu\nu}={\rm diag}(+--)$. In turn, unitarity reflects the probability conservation, which means that ghosts, whose one-particle norm states are negative, do not show up in the spectrum,  or they decouple, {\it i.e.} the one-particle norm states that refer to physical quantities are all non-negative.Since the model analyzed here has the poles of the bosonic propagators (\ref{prop1})--(\ref{prop3}) depending on the fermion flavor numbers ($n_\psi$, $n_\chi$ and $n_\o$) and coupling constants ($e$ and $g$), it is also worth checking whether some conditions involving them are necessary to assure causality and unitarity at the tree-level.

In what concerns causality, each fermion propagator (\ref{prop4}) has a pole $k^2=m^2>0$, which is positive definite, thus the three fermions are bradyons. On the other hand, the boson propagators (\ref{prop1})--(\ref{prop3}) exhibit three poles: 
\begin{equation}\label{poles1}
k^2=0~,~~ k^2= \mu_{\pm}^2=\frac{1}{2}\left(E^2+G^2+2H^2\pm\sqrt{(E^2-G^2)^2+4H^2(E+G)^2}\right)~,
\end{equation}
where causality is satisfied only if $k^2=\mu_{\pm}^2\geq 0$, with 
$\mathcal{D}=(k^2-\mu_+^2)(k^2-\mu_-^2)$. However, it can be checked that, from (\ref{definitions}) and (\ref{poles1}), whatever the values of $E$, $G$ and $H$, all the poles are always real and non-negative, thus constraints on the flavors $n_\psi$, $n_\chi$ and $n_\o$, and couplings constants, $e$ and $g$, are not necessary to ensure causality at the tree-level.

Concerning the unitarity, thus the absence of negative norm states in the spectrum, so as to verify it, we couple the propagators to external currents, and by means of current-current transition amplitudes,  
$\ca_{\phi_1 \phi_2}\equiv J^\m_{\phi_{1}}\D_{\m\n}^{\phi_1 \phi_2}J^\n_{\phi_2}$, we  calculate the imaginary part of the residues corresponding to each propagator pole \cite{nieuwenhuizen}. If for any transition amplitude, ${\mathcal Im}\{{\mathcal Res}\{\ca_{\phi_1 \phi_2}|_{\rm pole}\}\}$ is negative, unitarity is spoiled,
thus there are negative norm states (ghosts) in the spectrum. 

We start from analyzing the fermionic spectrum through the fermion-fermion amplitude. The spinor currents $J_\phi$ and $\bar{J_\phi}$, where $\phi=\{\psi,\chi,\o\}$, are given by $J_\phi=(\theta_1,\theta_2)^{\intercal}$ and $\bar{J}_\phi=J^\dagger_\phi \g_0=(\theta_1^*,-\theta_2^*)$. However, due to the fact that the fermions are massive, we can choose the rest frame $k^\m=(m,0,0)$. So, it follows that:
\begin{equation} \label{fermion}
  {\mathcal Im}\{{\mathcal Res}\{\mathcal{A}_{\phi\phi}|_{k^2=m^2}\}\}=
  {\mathcal Im}\left\{\lim_{k^2\to m^2}(k^2-m^2)i\bar{J}^\dagger_\phi\frac{{\sl k}-n_\phi/|n_\phi|m}{k^2-m^2}J_\phi\right\}=m(|\theta_1|^2+|\theta_2|^2)-\frac{n_\phi}{|n_\phi|}m(|\theta_1|^2-|\theta_2|^2)~,
\end{equation}
which is always positive definite since $n_\phi/|n_\phi|=\pm 1$, consequently there are neither ghosts nor tachyons, so unitarity and causality are guaranteed in the fermionic spectrum. Also, from (\ref{fermion}) we see that the three fermions propagate two degrees of freedom with mass $|m|$.

In the following, let us analyze the bosonic spectrum. The vector currents $J_A^\m$ and $J_Z^\m$ associated to the fields $A^\m$ and $Z^\m$, respectively, can be decomposed into a 1+2 Minkowski space-time basis, {\it i.e.} $k^\m=(k^0,k^1,k^2)$, $\tilde{k}^\m=(k^0,-k^1,-k^2)$ and $\ve^\m=(0,\ve^1,\ve^2)$, satisfying the covariant constraints $k^\m \ve_\m=\tilde{k}^\m \ve_\m=0$, $k^\m k_\m=\tilde{k}^\m\tilde{k}_\m=\m^2$ and $\ve^\m \ve_\m=-1$,
where $\m$ is the mass of some vector gauge field degree of freedom, that might be zero as well. Furthermore, since the currents are conserved, they are constrained by the continuity equations, $k_\m J_A^\m=k_\m J_Z^\m=0$. For instance, firstly taking into consideration the massive pole $\mu_+^2$, we choose the rest frame by setting $k^\m=\tilde{k}^\m=(\mu_+,0,0)$. From the covariant constraints and current conservation, the vector currents expressed in terms of the space-time basis, $J_A^\m=A_A k^\m+B_A\tilde{k}^\m + C_A\ve^\m$ and $J_Z^\m=A_Z k^\m+B_Z\tilde{k}^\m + C_Z\ve^\m$, fulfill the conditions $k_\m J_A^\m=(A_A +B_A)\mu_+^2=0$ and $k_\m J_Z^\m=(A_Z +B_Z)\mu_+^2=0$, implying that $A_A=-B_A$ and $A_Z=-B_Z$. Moreover, in the rest frame $k^\m=\tilde{k}^\m$, then $J_A^\m= C_A\ve^\m$ and $J_Z^\m= C_Z\ve^\m$. Analogously, for the massive pole $\mu_-^2$, $J_A^\m=C'_A\ve^\m$ and $J_Z^\m=C'_Z\ve^\m$. The current-current transition amplitudes of the gauge fields read
\begin{equation}
	J^{\m}_{A} \D^{AA}_{\m\n} J^{\n}_{A} = i \frac{k^2-G^2-H^2}{(k^2-\m_+^2)(k^2-\m_-^2)} C_A^2~,~~~~~~J^{\m}_{Z} \D^{ZZ}_{\m\n} J^{\n}_{Z} = i \frac{k^2-E^2-H^2}{(k^2-\m_+^2)(k^2-\m_-^2)} C_Z^2~.
\end{equation}
In \cite{kaplan_2020}, the photon is considered as the field $A_\m$, in the limit where $g \gg e$.
In this limit, $\m_+ \approx G^2+H^2$ and $\m_- \approx E^2+H^2$, resulting in
a squared mass $E^2+H^2$ for the photon, and it increases with $g^2$, \textit{i.e.} the mass of the photon increases with $g$. On the other hand, $Z_\m$ acquires a squared mass $G^2+H^2$, which increases with $g^4$. Consequently,
in this limit, the field $Z_\m$ is much heavier than $A_\m$, and it can be integrated
out, leading to the derivation of the effective action for the photon, as mentioned in \cite{kaplan_2020}. However, the field $A_\m$ is not the true photon, as will be discussed in the following when we present the genuine one, which
is much lighter than $A_\m$, considering $g \gg e$.

At this moment, it is opportune to verify if whether or not there exists spurious crossover propagating degrees of freedom, namely, if the gauge fields $A_\m$ and $Z_\m$ are actually the fundamental fields. This issue can be achieved by analyzing the current-current transition amplitude for the mixed propagator (\ref{prop3}). To do so, we consider the pole $\mu^2_+$ and compute the imaginary part of the residue of the mixed current-current amplitude at that pole:
\begin{align}\label{mixed1}
  \nonumber
  {\mathcal Im}\{{\mathcal Res}\{\ca_{AZ}|_{k^2=\mu_+^2}\}\}&={\mathcal Im}\left\{ 
  \lim_{k^2 \to \mu_+^2}  \left[
  (k^2-\mu_+^2)i\frac{H (E+G)}{(k^2-\mu_+^2)(k^2-\mu_-^2)}J_Z^\m \left(\eta_{\m\n}-\frac{k_\m k_\n}{k^2}\right)J_A^\n\right]\right\}\\
  &=-C_A C_Z\frac{H(E+G)}{\mu_+^2-\mu_-^2}~.
\end{align}
Meanwhile, for the pole $\mu^2_-$:
\begin{equation}\label{mixed2}
{\mathcal Im}\{{\mathcal Res}\{\ca_{AZ}|_{k^2=\mu^2_-}\}\}=C'_A C'_Z\frac{H(E+G)}{\mu_+^2-\mu_-^2}~.
\end{equation}
It should be stressed that the previous transition amplitudes (\ref{mixed1}) and (\ref{mixed2}) are not necessarily zero, indicating therefore that $A_\m$ and $Z_\m$ are not the fundamental fields. Besides that, the amplitudes (\ref{mixed1}) and (\ref{mixed2}) might be negative, whatever the values of $C_A$, $C_Z$, $C'_A$ and $C'_Z$, jeopardizing  the spectrum consistency through unitarity violation in the bosonic sector, thanks to propagating negative norm states. In conclusion, owing to the fact that the gauge fields, $A_\m$ and $Z_\m$, are not the fundamental fields indeed, and bearing in mind that those vector fields carry ghosts as propagating degrees of freedom, consequently harming the spectrum, we ought to find out the fundamental fields by diagonalizing the gauge field action $\S_{AZ}$ (\ref{bosonicact2}).

\textit{The fundamental gauge fields}---. In order to perform an appropriate analysis concerning the tree-level unitarity of the model, the fundamental gauge fields have to be determined as linear combinations of the original vector fields, $A_\m$ and $Z_\m$. To do so, the gauge field action to be diagonalized is taken as: 
\begin{equation}\label{AZ0}
  \widetilde{\S}_{AZ}=\int d^3x~ V^\intercal {\cal O} V
  ~,~~
  V\equiv\begin{bmatrix}
   A_\alpha\\
   Z_{\alpha}
  \end{bmatrix}~,
  ~~
   {\cal O}\equiv\begin{bmatrix}
   \dfrac{1}{2}(\square\Theta^{\mu\nu} +ES^{\mu\nu}) & \dfrac{1}{2}HS^{\mu\nu}\\
   \dfrac{1}{2}HS^{\mu\nu}  & \dfrac{1}{2}(\square \Theta^{\mu\nu} +GS^{\mu\nu})
  \end{bmatrix}~.
\end{equation}
Since ${\cal O}$ is a real symmetric matrix, hence diagonalizable, this allows the decoupling of the gauge fields, $A_\alpha$ and $Z_\alpha$, suppressing the mixed Chern-Simons term -- as discussed above, it is the term in the action $\S_{AZ}$ (\ref{bosonicact2}) that surely induces the emergence of negative norm states in the spectrum. At this time, for the purpose of diagonalize the wave operator ${\cal O}$ we solve the eigenvalue problem by finding the operator $\Lambda$ such that ${\cal O}=diag(\Lambda_1^{\mu\nu},\Lambda_2^{\mu\nu})$, then the two eigenvalues read
\begin{equation} \label{eigenvalues}
  \Lambda_1^{\mu\nu}=\square \Theta^{\mu\nu} + \frac{1}{2}\left(G+E+\sqrt{(G-E)^2+4H^2}\right)S^{\mu\nu}~,~~
  \Lambda_2^{\mu\nu}=\square \Theta^{\mu\nu} + \frac{1}{2}\left(G+E-\sqrt{(G-E)^2+4H^2}\right)S^{\mu\nu}~,
\end{equation}
with the orthonormal eigenvectors being $v_{\Lambda_1}=(-\xi,\zeta)$ and $v_{\Lambda_2}=(\zeta,\xi)$, where
\begin{equation} \label{eigenvectors}
  \xi=\frac{2H}{\sqrt{4H^2+\left[(E-G)-\sqrt{(E-G)^2+4H^2}\right]^2}}~,~~
  \zeta=\frac{(E-G)-\sqrt{(E-G)^2+4H^2}}{\sqrt{4H^2+\left[(E-G)-\sqrt{(E-G)^2+4H^2}\right]^2}}~.
\end{equation}
Consequently, the fundamental gauge fields, $W_+^\mu$ and $W_-^\mu$, are given by 
\begin{equation} \label{w+w-}
  %%%%%%%%%%%%%%%%%%%%
  \begin{bmatrix}
    W_+^\mu\\
    W_-^\mu
  \end{bmatrix}
  %%%%%%%%%%%%%%%%%%%%
  =
  %%%%%%%%%%%%%%%%%%%%
  \begin{bmatrix}
    -\xi & \zeta\\
     \zeta & \xi
  \end{bmatrix}
  %%%%%%%%%%%%%%%%%%%%
  \begin{bmatrix}
    A^\mu\\
    Z^\mu
  \end{bmatrix}~.
\end{equation}
In the sequence, by taking into account (\ref{eigenvalues}), (\ref{eigenvectors}) and (\ref{w+w-}), and rewriting the action $\S_{AZ}$ (\ref{bosonicact2}) in terms of the fields $W_+^\mu$ and $W_-^\mu$, it follows that
\begin{align} \label{w+w-action}
&\S_{W_\pm} =\int d^3x\bigg\{ \frac{1}{2}W_+^\m\bigg[\square \left(\Theta_{\mu\nu}+\frac{1}{\a_+}\O_{\m\n}\right)+M_+ S_{\mu\nu} \bigg]W_+^\n + \frac{1}{2}W_-^\m\bigg[\square \left(\Theta_{\mu\nu}+\frac{1}{\a_-}\O_{\m\n}\right)+M_- S_{\mu\nu} \bigg]W_-^\n\bigg\}\\
  &= \int d^3x \bigg\{-\frac{1}{4}F_+^{\m\n}F^+_{\m\n}-\frac{1}{4}F_-^{\m\n}F^-_{\m\n} + 
  \frac{1}{2}M_+ \ve_{\m\r\n}W_+^\m \partial^\r W_+^\n + \frac{1}{2}M_- \ve_{\m\r\n}W_-^\m \partial^\r W_-^\n -  
  \frac{1}{2\a_+}(\pa_\m W_+^\m)^2 - \frac{1}{2\a_-}(\pa_\m W_-^\m)^2 \bigg\}~,\nonumber
\end{align}
where $F^+_{\m\n}$ and $F^-_{\m\n}$ are the field strengths associated to $W_+^\mu$ and $W_-^\mu$, respectively, with  $M_+=\xi^2E+\zeta^2G-2\xi\zeta H$ and $M_-=\zeta^2E+\xi^2G+2\xi\zeta H$. Recalling the regime $g \gg e$ discussed in \cite{kaplan_2020}, hence $|M_+| \gg |M_-|$, it results that $\xi\approx H/G$ and $\zeta\approx -1$, as a consequence, $M_+\approx G$ is the $W$-boson ($W_+$) mass, whereas $M_-\approx E-H^2/G$ is the photon ($W_-$) mass. Also, reminding the definitions (\ref{definitions}), it follows that:
\begin{equation}
M_+\approx 2g^2(n_\chi+n_\o) ~,~~ M_-\approx 2e^2\left(n_\psi + \frac{n_\chi n_\o}{n_\chi + n_\o}\right)~.
\end{equation} 
where the results of \cite{kaplan_2020} is obtained up to a redefinition, $e^2\mapsto e^2/(4\pi)$ and $g^2\mapsto g^2/(4\pi)$. Beyond that, from action (\ref{w+w-action}) by integrating out the heavier gauge field ($W_+$), we get the effective action for the photon ($W_-$):
\begin{equation} \label{photon}
\S_{\rm photon} = \int d^3x \bigg\{-\frac{1}{4}F_-^{\m\n}F^-_{\m\n} + \nu e^2 \ve_{\m\r\n}W_-^\m \partial^\r W_-^\n \bigg\} + \cdots ~,~~ 
\nu=\left(n_\psi + \frac{n_\chi n_\o}{n_\chi + n_\o}\right)~.
\end{equation}
Accordingly, it should be stressed that the action (\ref{w+w-action}), expressed by means of the vector fields $W^+_\m$ and $W^-_\m$, recovers the phenomenology and the Hall conductivity presented \cite{kaplan_2020}. Furthermore, as shall be presented below, it circumvents the issue of ghost states in the spectrum which would spoil, like previously analyzed, the unitarity of the model even at the tree-level. We also emphasize that the mass of $W^-_\m$, given by $M_-$, is independent of $g$ when considering the limit $g\gg e$, and in this situation, $W^-_\m$ is much lighter than $A_\m$, as the mass of the latter increases with $g$. This is due to the fact that $A_\m$ has also a contribution of $W^+_\m$, although small, in the considered limit. This distinction significantly impacts the spectrum of the photon in the effective action.

At this point, let us consider the action (\ref{w+w-action}) expressed in terms of the fundamental gauge fields, $W^+_\m$ and $W^-_\m$, in order to calculate the propagators $\D^{++}_{\m\n}$ and $\D^{--}_{\m\n}$ related to them. Thereupon, we get
\begin{equation}\label{newprop1}
  \D^{++}_{\m\n}(k)=-\frac{i}{k^2-M_+^2}\left(\eta_{\m\n}-\frac{k_\m k_\n}{k^2}\right)-i \frac{\a_+}{k^2}\frac{k_\m k_\n}{k^2}+\frac{M_+}{k^2(k^2-M_+^2)}\ve_{\m\r\n}k^\r~,
\end{equation}
\begin{equation}\label{newprop2}
  \D^{--}_{\m\n}(k)=-\frac{i}{k^2-M_-^2}\left(\eta_{\m\n}-\frac{k_\m k_\n}{k^2}\right)-i \frac{\a_-}{k^2}\frac{k_\m k_\n}{k^2}+\frac{M_-}{k^2(k^2-M_-^2)}\ve_{\m\r\n}k^\r~,
\end{equation}
where $M_+$ is the mass parameter of $W_+$-field and $M_-$ the one associated to $W_-$-field. It is appropriate to mention here that subsequently to diagonalization no condition has to be imposed on the flavors, $n_\psi$, $n_\chi$ and $n_\o$, as well as upon the coupling constants, $e$ and $g$, in order to assure the causality of the model, since $M_+^2$ and $M_-^2$ are always real and positive definite, which confirms our previous analysis that the model is causal and no constraints on the fermion flavor numbers and coupling constants are necessary. However, it lacks the final analysis about unitarity and propagating degrees of freedom. 

Analogously to what already discussed for the undiagonalized $Z_\mu$ and $A_\mu$ propagators (\ref{prop1})--(\ref{prop3}), let us now take into account the diagonalized $W^+_\m$ and $W^-_\m$ propagators, (\ref{newprop1}) and (\ref{newprop2}). The conserved currents coupled to the propagators $\D^{++}_{\m\n}$ and $\D^{--}_{\m\n}$ are decomposed as $J_+^\m= A_+ k^\m+B_+ \tilde{k}^\m+C_+\ve^\m$ and $J_-^\m= A_- k^\m+B_- \tilde{k}^\m+C_-\ve^\m$, fulfilling $k_\m J_+^\m=k_\m J_-^\m=0$. Bearing in mind the massive poles, $M_+^2$ and $M_-^2$, hence $J^\m_{+}=C_+ \ve^\m$ and $J^\m_{-}=C_- \ve^\m$, consequently the imaginary part of the current-current transition amplitudes are as follows: 
\begin{equation} \label{massive}
  {\mathcal Im}\{{\mathcal Res}\{\ca_{++}|_{k^2=M_+^2}\}\}=-J_+^\m \eta_{\m\n} J_+^\n=C_+^2>0~,~~
  {\mathcal Im}\{{\mathcal Res}\{\ca_{--}|_{k^2=M_-^2}\}\}=-J_-^\m \eta_{\m\n} J_-^\n=C_-^2>0~,
\end{equation}
which can be concluded that the both gauge fields $W^+_\m$ and $W^-_\m$ carry each one two propagating degrees of freedom with masses $|M_+|$ and $|M_-|$, respectively. To end the unitarity analysis it remains to consider the massless pole. However, assuming $k^\m=(\m,0,\m)$, it is straightforwardly verified that, since $J_+^\m= (\m A_+,C_+,\m A_+)$ and $J_-^\m= (\m A_-,C_-,\m A_-)$: 
\begin{equation} \label{massless}
  {\mathcal Im}\{{\mathcal Res}\{\ca_{++}|_{k^2=0}\}\}=0~,~~{\mathcal Im}\{{\mathcal Res}\{\ca_{--}|_{k^2=0}\}\}=0~.
\end{equation}
Lastly, we conclude that the diagonalized model, {\it i.e.} the gauge sector action (\ref{w+w-action}) of the Kaplan-Sen model expressed in terms of the fundamental vector fields, $W^+_\m$ and $W^-_\m$, is also unitary at tree-level, as well as no restrictions on the coupling constants and fermion flavors are required at all.

\textit{Conclusions}---. In summary, it has been proved that the model presented here, defined by the fermion piece of the  action (\ref{action}) -- expressed by the three family fermions, $\psi_i$, $\chi_j$ and $\o_k$ -- added by the gauge action (\ref{w+w-action}) --  which is written in terms of the fundamental vector fields, $W^+_\m$ and $W^-_\m$ -- is causal and unitary at the tree-level. Moreover, the spectrum consistency does not demand any necessary condition to be fulfilled by the coupling constants, $e$ and $g$, and the fermion flavors, $n_\psi$, $n_\chi$ and $n_\o$. It should be stressed, however, that the unitarity at the tree-level was only guaranteed after the diagonalization procedure and the determination of the fundamental vector fields. Concerning the degrees of freedom of the propagating quanta, each of the three fermions propagates two degrees of freedom with mass $|m|$, while both gauge fields exhibit two degrees of freedom as well, with masses $|M_+|$ and $|M_-|$. Additionally, we have shown that if $W_+$ is much heavier than $W_-$, thus whenever $g \gg e$, the action (\ref{w+w-action}) gives rise, by integrating out the $W^+_\m$-field, to the effective action (\ref{photon}) where the field $W^-_\m$ plays the role of the photon, exhibiting precisely the phenomenology discussed by Kaplan and Sen in \cite{kaplan_2020}. As an outlook, it should be interesting to analyze this model, now expressed in terms of the fundamental fields, in the algebraic renormalization framework at all orders in perturbation theory, since the model might be ultraviolet finiteness, namely quantum scale invariant.

\textit{Acknowledgements}---. The authors thank J.A. Helay\"el-Neto for the helpful discussions. CAPES-Brazil is acknowledged for invaluable financial help.

\end{document}